\documentclass[12pt]{article}
\usepackage{amssymb,amsmath,epsfig}
\begin{document}
\parindent=0pt
\parskip=6pt
\rm
\begin{center}
\Large\textbf{Phenomenological description of anisotropy effects  in some ferromagnetic superconductors}
\medskip

\normalsize
\textbf{Diana V. Shopova$^\dag$\footnote{Corresponding author: sho@issp.bas.bg}, Michail D. Todorov$^\ddag$}
\medskip\small

\textit{$^\dag$TCCM Research Group, Institute of Solid State Physics,}
 \emph{Bulgarian Academy of Sciences, BG-1784 Sofia, Bulgaria}\smallskip

\textit{$^\ddag$Department of Applied Mathematics and Computer Science,}
 \emph{Technical University of Sofia, 1000 Sofia, Bulgaria}\\
\end{center}


\begin{abstract}
We study phenomenologically by using the previously derived Landau free energy, the role of anisotropy  in ferromagnetic superconductors UGe$_2$, URhGe, and UCoGe.  The three compounds are separately discussed with the special stress on UGe$_2$. The main effect comes from the strong uniaxial anisotropy of magnetization while the anisotropy of Cooper pairs and crystal anisotropy only slightly change the phase diagram in the vicinity of Curie temperature. The limitations of this approach are also discussed.
\end{abstract}

\normalsize

\section{\label{1} Introduction}
The discovery of ferromagnetic superconductors  UGe$_2$~\cite{Saxena:2000}, URuGe~\cite{Aoki:2001}, UCoGe~\cite{Huy:2007}, in which ferromagnetic ordering coexists with superconductivity, has given a new trend in understanding of unconventional superconductivity. The pressure-temperature phase diagrams of these compounds differ, but the common feature is that the superconductivity occurs in the domain of ferromagnetic phase and the superconducting transition temperature $T_s$ is lower than the Curie temperature, $T_c$. ${\rm UGe}_2$ orders ferromagnetically at relatively high Curie temperature of 53 K and superconductivity appears upon the application of pressure  of about 1 GPa, and at low temperature $< 1$  K. The increase of pressure to the critical value $P_c = 1.5 $ GPa results in disappearance of both ferromagnetic and superconducting order. \\ URuGe and  UCoGe are weaker ferromagnets with $T_c$ of 9.5 K and 3 K, respectively and the superconducting phase appears at ambient pressure as well. For URuGe the increase of pressure leads to the collapse of superconductivity at about 4 GPa, while for UCoGe the phase transition line gradually grows reaching maximum at 1.1 GPa, where the ferromagnetic order collapses and superconductivity persists also in the paramagnetic region. All three uranium compounds have orthorhombic crystal structure  with highly anisotropic magnetic moment of Ising type. For detailed presentation of ferromagnetic superconductors, see, for example, the recent review~\cite{Aoki:2012}.\\
It is commonly accepted that 5f electrons of uranium atoms are responsible for both ferromagnetic and superconducting orders. In the presence of magnetization, the ferromagnetic exchange field is expected to rule out spin-singlet Cooper pairing and unconventional superconductivity of p-type, mediated through some magnetic mechanism is considered as the most likely. The experimental discovery of huge upper critical field in URhGe and UCoGe also confirms the triplet pairing because the Pauli paramagnetic effect characteristic of spin singlet pairing is absent there, see, for example~\cite{Aoki:2009} and the papers cited therein.
The anisotropic properties of superconductivity in these compounds are vastly studied experimentally, especially the anisotropic properties of upper critical field~\cite{Aoki:2012}.\\
Here we will study phenomenologically the role of magnetic, crystal and Cooper-pair anisotropy on the phase diagram and possible phases using the previously derived Ginzburg-Landau free energy~\cite{Cottam:2008, Shopova:2009}.

\section{\label{sec:level1}Landau free energy}

We will  consider only the Meissner phases -- pure superconductors and phases of coexistence of ferromagnetism and superconductivity in the absence of external magnetic field.  In earlier papers~\cite{Cottam:2008, Shopova:2009} we did not consider the Cooper-pair and crystal anisotropy as the main purpose was to describe the $P-T$ phase diagram. Because the coexisting phase of ${\rm UGe}_2$ is totally within the domain of ferromagnetic phase we have assumed that it is the presence of ferromagnetism that triggers the appearance of superconductivity under external pressure. The pressure  participates only through the linear dependence of Curie temperature on $P$, namely, $T_c=T_{c0}(1-P/P_0)$, where $T_{c0}$ is the Curie temperature at zero (ambient) pressure and $P_0$ is the pressure close to the critical $P_c$ where ferromagnetism and superconductivity disappear.  This free energy and the obtained results may be a good starting point for the description of $(P,T)$ phase diagram in ferromagnetic superconductors.

The general form of Ginzburg-Landau free energy for the Meissner phase of ferromagnetic superconductors with $p$-pairing is:

\begin{equation}
\label{Eq1} F(\vec{M},\vec{\psi}) = F_M + F_{sc} + F_{int}.
\end{equation}
\noindent
The free energy density $F_{sc}$ of pure superconducting system is expanded up to the fourth order in superconducting order parameter, including the respective anisotropic terms. Here we suppose tetragonal symmetry for superconductors with  triplet Cooper pairing~\cite{Volovik:1985}. Although all three uranium compounds UGe$_2$, URhGe and UCoGe  have orthorhombic symmetry and the structure of superconducting order parameter for orthorhombic symmetry has been derived by general group considerations~\cite{Fomin:2001, Mineev:2002}, here we shall not consider for the time being the anisotropy in $(x,y)$ plane, but only the uniaxial anisotropy, connected with the Ising-like anisotropy of magnetization, \emph{i.e.,}
\begin{equation}
\label{Eq2} F_{sc}=a_s |\overrightarrow{\psi}|^2 + \frac{b_s}{2} |\overrightarrow{\psi}|^4 +\frac{u_s}{2} |\overrightarrow{\psi}^2|^2 + \frac{v_s}{2}(|\psi_1|^4+|\psi_2|^4),
\end{equation}
with $\overrightarrow{\psi}=(\psi_1, \psi_2, \psi_3)$ -- complex three-component superconducting order parameter. The Landau material parameters are given by $a_s=\alpha_s[T-T_s(P)]$ , where $T_s(P)$ is the critical temperature for pure superconducting system and $b_s >0$. The terms with $u_s$ and $v_s$ represent the anisotropy of spin-triplet Cooper pairs and uniaxial crystal anisotropy, respectively.\\
The strong uniaxial anisotropy of magnetic moment plays an important role for proposed magnetic mediated mechanisms for the appearance of triplet pairing  favored by longitudinal magnetic fluctuations, see, for example~\cite{Hattori:2012}.\\
 The ferromagnetic energy density up to the forth order in magnetization $\overrightarrow{M}$ is denoted by $F_M$
 \begin{equation}\label{Eq3}
 F_M= a_f |\overrightarrow{M}|^2 + \frac{b_f}{2} |\overrightarrow{M}|^4.
  \end{equation}
  Here $a_f=\alpha_f[T^n-T^n_f(P)]$; $n=1$ gives the usual Landau form for $a_f$, and $n=2$ describes the spin fluctuation theory~\cite{Yamada:1993}; $b_f>0$.\\
The interaction between the superconducting and magnetic order parameters is given by $F_{int}$ :
  \begin{equation}\label{Eq4}
F_{int} = i\gamma_0 \overrightarrow{M}.(\overrightarrow{\psi}\times \overrightarrow{\psi}^\ast) + \delta \overrightarrow{M}^2|\overrightarrow{\psi}|^2,
  \end{equation}
with $\gamma_0 \sim J$, where $J$ is the ferromagnetic exchange constant.

The choice of uniaxial magnetic anisotropy means that magnetic moment  in the above equations can be represented in the form $\overrightarrow{M} =(0,0,M_z)$ by choosing $z$ as the easy axis of magnetization.

To facilitate our considerations we make the free energy~(\ref{Eq1}) dimensionless.
To this end we introduce:
    \begin{equation}\label{Eq5}
        f=\frac{F}{b_f M_0^4},
    \end{equation}
    where $ M_0=\alpha_f T_{f0}/\sqrt{b_f}$ is the magnetic moment at $T=0$, $P=0$ with $T_{f0}$ the Curie temperature at zero pressure in the absence of superconducting order. The other parameters in~(\ref{Eq1}) become $t=(T-T_f)/T_{f0}$, \quad $r=\beta (T-T_s)/T_{f0}; \quad \beta = \alpha_{s}\sqrt{b_f/b}/\alpha_f$. Here we have introduced the notation $b=b_s+u_s+v_s$. The dimensionless order parameters are:
    $$m=\frac{M}{M_0}; \quad  \varphi_i=\phi_i e^{\theta_i}; \quad \phi_i=\frac{|\overrightarrow{\psi_i}|}{M_0 (b_f/b)^{1/4}}$$ and the parameters, describing interaction between superconducting order parameter and magnetic moment become: $\gamma=\gamma_0/(\sqrt{b}b_f^{1/4}); \quad  \gamma_1=\delta/\sqrt{b b_f}$.

    In this way the dimensionless free energy density takes the form
   \begin{multline}
    f = r(\phi_1^2+\phi_2^2+\phi_3^2)+\frac{1}{2}(\phi_1^2+\phi_2^2+\phi_3^2)^2\\ 
    -2w\left[\phi_1^2\phi_2^2\sin^2(\theta_2-\theta_1)+\phi_1^2\phi_3^2\sin^2(\theta_1-\theta_3)+\phi_2^2\phi_3^2\sin^2(\theta_2-\theta_3)\right]\\
    -v\phi_1^2\phi_2^2
      +2\gamma\phi_1\phi_2M\sin(\theta_2-\theta_1)+\gamma_1(\phi_1^2+\phi_2^2+\phi_3^2)M^2+tM^2+\frac{1}{2}M^4.
   \label{Eq6}
   \end{multline}
The parameters of Cooper-pair anisotropy $w=w_s/b$ and uniaxial crystal anisotropy $v=v_s/b$ can take both positive and negative values, but their modulus remains smaller than unity by this definitions.\\
     The equilibrium phases are found from the equations of state
     \begin{equation}\label{Eq7}
       \frac{ \partial f(x_i)}{\partial x_i}=0,
    \end{equation}
    with $\{x_i\}=(m, \phi_1, \phi_2, \phi_3; \theta_1,\theta_2,\theta_3).$
    The stability of obtained phases and phase transition lines can be calculated from the stability matrix, with elements defined by:
    \begin{equation}\label{Eq8}
      A_{ij}= \frac{ \partial^2 f}{\partial x_i \partial x_j},
    \end{equation}
The above expression~\eqref{Eq6} for the free energy is not so simple  and there is a great number of  solutions for  the equations of state~(\ref{Eq7}). Here we focus on those new solutions that appear because of taking account of both the anisotropy of Cooper pairs and crystal anisotropy. The stability in case of too many phases cannot be calculated only trough the stability matrix~(\ref{Eq8}) but in order to find the phases corresponding to a global minimum of the free energy, direct comparison of respective free energies should be done.

   The calculations are accomplished in the space of parameters $r(T,P)$, $t(T,P)$; $T$ and $P$ are the temperature and pressure, respectively. Here we do not make any assumptions for the particular dependence of $r$ and $t$ on the pressure, as we want to see the general trend and possibility to use the free energy~(\ref{Eq6}) for qualitative estimates. In order to describe the $(P,T)$ phase diagrams and make quantitative comparison with the available experimental data, the $(P,T)$ diagram for UGe$_2$, URhGe, and UCoGe should be separately considered as discussed below.

   \section{\label{sec:leve8} Results and Discussion}
   The crystal anisotropy and the Cooper-pair anisotropy lift the degeneracy of the free energy~(\ref{Eq6}) and change both the number and stability domains of the possible phases. Here we assume that $\gamma_1 > 0$ as the negative sign means some redefinition of the free energy in order to ensure it is limited at infinity.
   The basic solution that gives a ferromagnetic superconducting phase is:
   \begin{equation}\label{Eq9}
   \phi_1=\phi_2=\phi,\ \phi_3=0, \ m\neq 0, \ \cos(\theta_1-\theta_2)=0.
   \end{equation}
   This is a two-domain phase with mutually perpendicular components of the superconducting order parameter in the complex plane. The domains differ in the sign of magnetic moment and $\sin(\theta_1-\theta_2)$  but in this approximation the domains are undistinguishable with equal energies and the inclusion of Cooper-pair and crystal anisotropy terms does not lift this degeneracy. Here we consider only the positive sign of $m$.

    To be more explicit, we write the equations of state for superconducting order parameter and magnetization:
   \begin{equation}\label{Eq10}
   \phi\left[r+\phi^2(2-2w-v)-\gamma m+\gamma_1 m^2\right] = 0,
\end{equation}
and
   \begin{equation}\label{Eq11}
   tm +m^3+2\gamma_1 m \phi^2 -\gamma \phi^2 =0.
   \end{equation}
   Solving together the above equations we find the equilibrium value of $ r(m,t)=r_0$, for which the ferromagnetic superconductor~(\ref{Eq9}) exists:
 \begin{equation}\label{Eq12}
   r_0=\frac{-2m}{m_0-m} \left[(1-\gamma_1^2-w-v/2)m^2+\frac{3}{2}\gamma_1\gamma m- (1-w-v/2) t +\frac{\gamma^2}{2}\right].
   \end{equation}
   The value $m_0$ of magnetic moment corresponds to the maximum of phase line  $r(t)$ for the transition from ferromagnetic to coexisting phase, see, for example Fig.~1 below, at $t=-\gamma^2/(4 \gamma_1^2)$.
   The expression for superconducting order parameter $\phi$ of coexisting phase becomes
\begin{equation}\label{Eq13}
    \phi^2=\frac{m (m^2+t)}{(m_0-m)}; \ \ m_0=\frac{\gamma}{2\gamma_1}.
\end{equation}
The superconducting order parameter is positive for $m<m_0, \ m>\sqrt{-t}$, and $m>m_0, \quad m<\sqrt{-t}$. So, there is a region of existence of $\phi$ also for $t>0$ in the paramagnetic region, where the transition from para phase to coexisting phase is of first order. The superconducting order parameter $\phi$ is determined mainly by the uniaxial magnetization $m$ and the interaction parameters $\gamma_1$ and $\gamma$, the influence of crystal and Cooper-pair anisotropy parameters $w$, $v$ within this approximation is manifested in the stability conditions.

The uniaxial ferromagnetic phase exists for $m^2=-t, \quad t<0$ and its domain of stability is given by $r\geq r_e$ with
\begin{equation}\label{Eq14}
   r_e=\gamma_1t+\gamma\sqrt{-t}
   \end{equation}
   and $r_e=r_0(m=\sqrt{-t})$.

    The crystal and Cooper-pair anisotropies  give rise to new coexisting phases: for  $\phi_1=\phi_2, \ \phi_3=0, \  m\neq 0$, one more coexisting phase is found with $\sin{(\theta_1-\theta_2)}=-\gamma m/(2 w \phi^2)$,  which is unstable. Another unstable ferromagnetic superconducting phase is given by
$\phi_1\neq \phi_2$, $\phi_3=0$, $m\neq 0$  and $2w+v <0$; this phase exists only for $r<0$. There is one more phase, namely, $\phi_1=\phi_2$, $\phi_3\neq0$, $m\neq0$, existing for $r>0$ with marginal stability; the phase with $\phi_1=\phi_2=\phi_3$ exists only for $v\equiv0$.

We show in Fig.~1 the phase diagram for the basic coexisting phase~(\ref{Eq9}).

\begin{figure}
\begin{center}
\epsfig{file=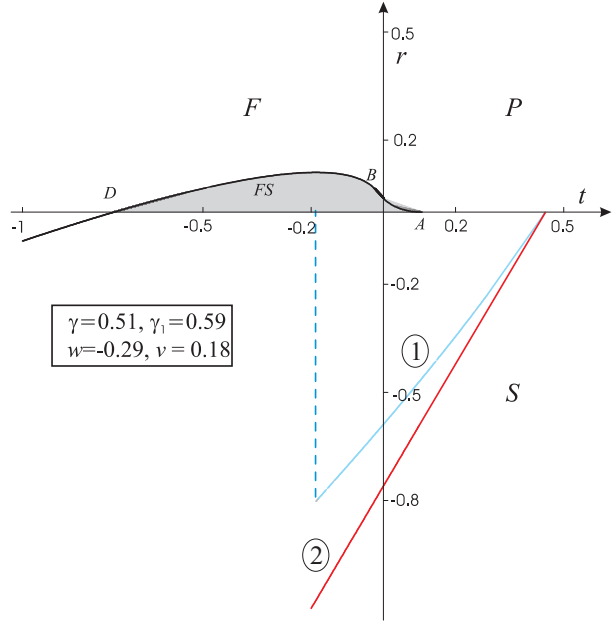,angle=-0, width=8cm}
\end{center}
\caption{\footnotesize The $(r,t)$ phase diagram of ferromagnetic superconductor: $P$ -- disordered phase; $F$ -- ferromagnetic phase; $FS$ -- ferromagnetic superconductor; S -- pure superconducting phases for $w<0, \ v>0$ } \label{Fig1}
\end{figure}
The area of coexisting phase is given in Fig.~1 in grey color. The  line $AB$ is first order line: for $t>0$ it describes the phase transition between the paramagnetic phase and the ferromagnetic superconductor, and for $t<0$ -- the transition between the ferromagnetic phase and coexisting phase. The point $B$ is tricritical one -- to the left of it, the phase transition between the ferromagnet and coexisting phase is of second order. The second order transition line is given by~(\ref{Eq14}). The effect of anisotropy parameters $w,\ v$ on the form of phase diagram is mainly on the first order phase transition lines in the interval $t_{A,B}=(t_A, t_B)$ given by
$$t_A=\frac{\gamma^2}{2(1-w-v/2)}, \qquad t_B=-\frac{\gamma^2(\gamma_1-\sqrt{1-w-v/2})^2}{4(1-\gamma1^2-w-v/2)^2}.$$
From the above expressions it is obvious that for positive $w$ and $v$, the interval $t_{A,B}$ becomes wider and the substantial effect is coming from the magnitude and sign of parameter $w$ of the Cooper pair anisotropy.  The negative sign of $w$ makes the interval $t_{A,B}$ smaller. For $w<0$ there is one more effect  as the existence and stability of the coexisting phase for $r<0$ is limited in the area enclosed by the curve numbered by 1 and the dashed vertical line. This dashed line marks the maximum on the phase line of transition between the coexisting and ferromagnetic phase, given by: $t_{\max}=-\gamma^2/(4\gamma_1^2)$ and $r_{\max}=\gamma^2/(4\gamma_1)$. For $t_D<t<t_{\max}$ with $t_D=-\gamma^2/\gamma_1^2$, the coexisting phase is stable only for $r>0$.

In Fig.~1 the phase diagram is shown for $w<0,\ v>0$. The stable pure superconducting phase for these parameters is defined by $\phi_1=\phi_2, \ \phi_3=0, \ m=0, \ \sin{(\theta_1-\theta_2)}=0$. This superconducting phase exists for $r<0$ and  is stable for $ r< (2-v)(2wt+\gamma^2)/(4 \gamma_1 w)$ - line 2.\\
The point $t_D > -1$ for the parameters in Fig.~1, which means that for $T=0, \ (t=-1)$ the only stable phase is ferromagnetic one. The ratio between the interaction parameters $(\gamma/\gamma_1)^2$ determines whether the ferromagnetic superconductor exists down to $T=0$. This is shown in Fig.~2 where $(\gamma/\gamma_1)^2 >1$, \emph{i.e.}, the linear interaction is stronger than the quadratic on, see~(\ref{Eq3}).

\begin{figure}
\begin{center}
\epsfig{file=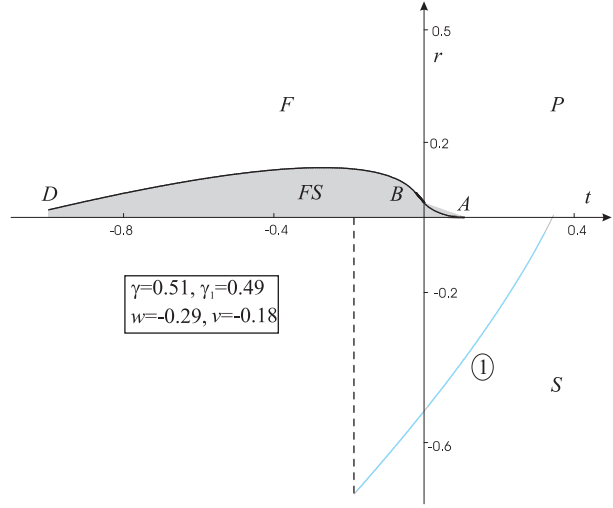,angle=-0, width=8cm}
\end{center}
\caption{\footnotesize The $(r,t)$ phase diagram of ferromagnetic superconductor for $\gamma >\gamma_1$ and $w<0 \ v<0$: $P$ -- disordered phase; $F$ -- ferromagnetic phase; $FS$ -- ferromagnetic superconductor; $S$ -- superconducting phases} \label{Fig2}
\end{figure}
The stability of coexisting phase for $r<0$, $w<0$ is limited, see the areas enclosed by line 1 and dashed line in Figs.~1 and 2, which is not the case for $w\geq 0$ where this is not limited in any way. To draw the whole phase diagram together with the pure superconducting phases, it is necessary to work in the $(P,T)$ space.\\
The superconducting phases that exist and are stable for $r<0$ and ($w<0,\ v<0$) are shown in Fig.~3.
\begin{figure}
\begin{center}
\epsfig{file=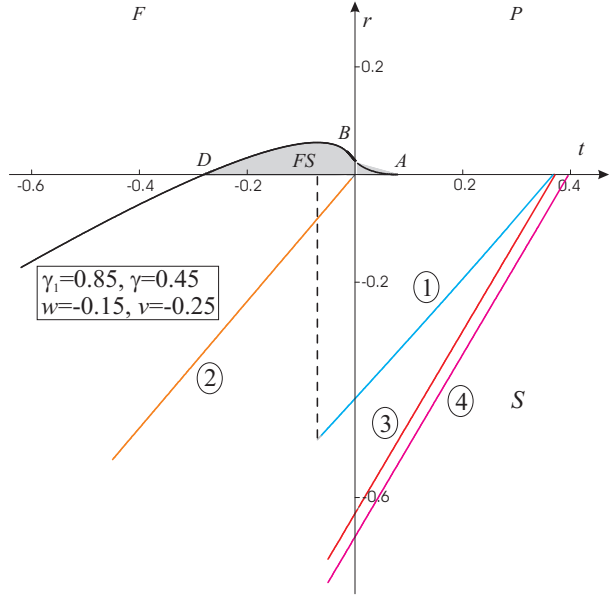,angle=-0, width=8cm}
\end{center}
\caption{\footnotesize The $(r,t)$ phase diagram for ferromagnetic superconductor for $w<0$, $v<0$ including the stability lines of purely superconducting phases: $P$ -- disordered phase; $F$ -- ferromagnetic phase; $FS$ -- ferromagnetic superconductor } \label{Fig3}
\end{figure}

The pure superconducting phases are given in Fig.~3 by their lines of stability numbered by 2, 3 and 4. The phase $\phi_2=\phi_3,\ m=0, \ \phi_1=0, \  \sin{(\theta_2-\theta_3)}=0$ exists for $r<0$ and is stable for $v <0, \  w<0$ and $r< r_2=t/\gamma_1$ with purely real collinear components, \emph{i.e.}, $\theta_
2=k\pi, \ \theta_3=l \pi$ . In  Fig.~3, $r_2$ is denoted by number 2. This phase may have also purely imaginary collinear components, namely, $\theta_2=(2k+1)\pi/2, \ \theta_3=(2l+1)\pi/2 $; the stability is defined by $w<0, 4w+v <0$, and in Fig.~3 it is shown under number 4 with $r<r_4=t/\gamma_1+2\gamma^2/\gamma_1(4w+v)$. Both phases have equal free energies but different domains of stability. The line 3 describes a pure superconducting phase with $\phi_2=\phi_1=0,\  m=0, \ \phi_3\neq 0$ which exists for $r<0$ and is stable for $\theta_3=(2k+1)\pi/2, \ w<0, \ r <r_3=t/\gamma_1 +\gamma^2/(2w \gamma_1)$.

The parameters in all figures above are only guiding to explain the different possibilities. Direct comparison with experiment requires that assumptions should be made about the dependence of Curie temperature on the pressure for each particular ferromagnetic superconductor UGe$_2$, URhGe and UCoGe. Then different  experimentally obtained phase diagrams and other experimental data should be used to find the material parameters of the Landau energy~(\ref{Eq6}) for each compound.\\
Here we discuss in detail the limitations for the use of free energy~(\ref{Eq6}) to the description especially of UGe$_2$. The calculations show that including the Cooper-pair and crystal anisotropy  terms for the description of $P-T$ phase diagram using the approach described in~\cite{Shopova:2009} with both linear, and quadratic dependence of Curie temperature on the pressure, does not rule out the main discrepancies between the experimentally found phase diagram and the one, calculated on the basis of Landau energy~(\ref{Eq1}).\\ First of all the transition from paramagnetic phase to ferromagnetic around $t=0$ experimentally is of first order up to the tricritical point $TCP$, where the transition becomes of second order down to $P=0$. See Fig.~4 for the experimental phase diagram of UGe$_2$.
\begin{figure}
\begin{center}
\epsfig{file=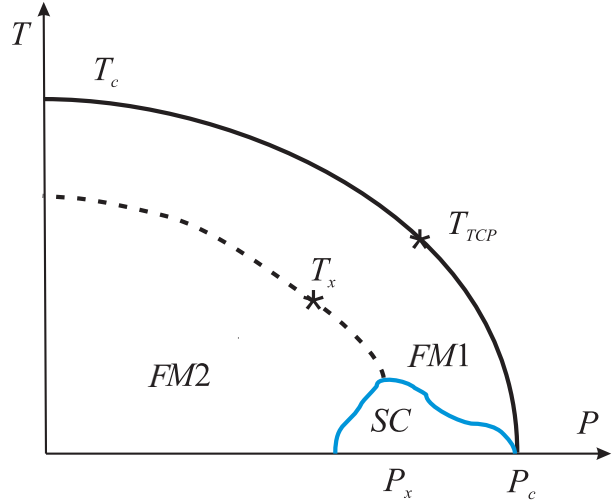,angle=-0, width=8cm}
\end{center}
\caption{\footnotesize An illustration of $T-P$ phase diagram
of UGe$_2$: $FM1$ -- high-pressure ferromagnetic phase, $FM2$ -- low-pressure ferromagnetic phase, $SC$-phase of coexistence of ferromagnetism and superconductivity.
$T_{x}(P)$ and $T_{c}(P)$ are the
respective magnetic phase transition lines; $TCP$ is tricritical point;
$P_c$ is the critical pressure, at which the ferromagnetic order disappears.} \label{Fig4}
\end{figure}
The first order phase transition from paramagnetic to $FM1$ phase can be modeled by expanding the magnetic energy density~(\ref{Eq3}) to $M^6$, then the parameter $b_f$ in front of $M^4$-term should change sign with pressure from negative to positive at tricritical point in Fig.~4. In~\cite{Shopova:2013} the first order transition from paramagnetic to low polarized ferromagnetic phase ($FM1$) is described in detail by including $M^6$-term and the coupling of elastic and magnetic degrees of freedom in~(\ref{Eq3}) and modeling the dependence of $b_f$ and $T_c$ on pressure.
The calculations show that when  the ferromagnetic energy density is expanded up to $M^6$ to determine the total phase diagram with the coexisting phase becomes too complicated as the
free energy density of the pure superconducting phase, (\ref{Eq2}) should be also expanded to sixth order in superconducting order parameter. Otherwise there is no region of existence of ferromagnetic superconductor to the left of maximum on the phase transition line from ferromagnetic to coexisting phase within the numerical accuracy of calculation. \\Taking into account the crystal and Cooper-pair anisotropy terms additionally complicates the problem as the Landau energy will depend on a bigger number of material parameters which are difficult to identify and compare with experimental data. Moreover, even the inclusion of sixth order term does not describe the transition from ferromagnetic low polarized ($FM1$) to high polarized ($FM2$) ferromagnetic phase, line $T_x$ in Fig.~4, which is important as the maximum at the phase line between ferromagnetic phase and coexisting phase is near to this ferromagnetic transition. It is clear from above considerations that direct application of Landau energy  is not suitable to use for UGe$_2$, as it is necessary to expand it to higher terms in $M$ -- sixth and eight  in order to describe the complex ferromagnetic transitions under pressure in this compound. As the appearance of superconductivity is triggered by the ferromagnetism this forces the inclusion of higher order terms also in the expansion of superconducting order parameter. The number of material parameters become too big in order to make any reasonable comparison with experiment. Another problem is related to the way, in which $T_c$ and $T_x$ depend on pressure and the compressibility. \\

The experimental $(P,T)$ phase diagram of  URhGe is also peculiar as the ferromagnetic transition line with the increase of pressure goes upward up to the highest pressure measured  of $\sim 13$  GPa. The magnetic moment exhibits also strong uniaxial anisotropy. The coexisting phase is completely within the ferromagnetic domain and occurs at ambient pressure, too. When pressure grows the coexisting phase line decreases and vanishes at $P=4$ GPa. The unusual experimental pressure dependence of Curie temperature cannot be easily modeled so to directly apply the Landau energy~(\ref{Eq1}) to the description of this phase diagram.

 UCoGe  is weak itinerant ferromagnet and is  the only uranium compound up to now, for which the superconductivity  appears not only in the domain of ferromagnetic phase but also in the paramagnetic region up to 2.2  GPa, while the ferromagnetic order is suppressed for $P\sim 1.1$  GPa. The ferromagnetic superconducting phase exists also at ambient pressure and magnetic moment shows uniaxial anisotropy of Ising type. In comparison with UGe$_2$ and URhGe, the phase diagram of UCoGe may be described using the Landau energy (\ref{Eq1}), if the real orthorhombic crystal anisotropy is considered together with the form of superconducting order parameter derived by general group considerations in~\cite{Fomin:2001, Mineev:2002} for this crystal symmetry.

 \section*{\label{sec:level1}Conclusion}
 In this paper we have analyzed phenomenologically the role of magnetic, Cooper-pair and crystal anisotropy for the description of phase diagrams of  some ferromagnetic superconductors. The phase diagram is mainly determined by the uniaxial anisotropy of magnetization; the Cooper-pair and crystal anisotropies slightly change the behavior of phase transitions lines in the vicinity of $r=0$, $t=0$. So the application of Ginzburg-Landau free energy up to the fourth order expansion of superconducting and magnetic order parameters may serve as initial estimate.
    The results and analysis show that the crucial point for the proper phenomenological description of phase of coexistence between ferromagnetism and superconductivity in UGe$_2$, URhGe and UCoGe is related to the ferromagnetic transitions under pressure in these compounds. More theoretical study is needed in this direction.

\end{document}